\documentclass[twoside]{article}
\usepackage{fleqn,espcrc2,epsf,rotate}

\usepackage{graphicx}

\newcommand{\AmS}{{\protect\the\textfont2
  A\kern-.1667em\lower.5ex\hbox{M}\kern-.125emS}}
\title{Microscopic Scenario for Striped Superconductors}

\author{J. Eroles\address{Theoretical Division, 
Los Alamos National Laboratory, Los Alamos, NM 87545}
                 \address{Centro At\'{o}mico Bariloche and Instituto Balseiro, 
S. C. de Bariloche, Argentina (8400)},
        G. Ortiz$^a$,
        A.V. Balatsky$^a$,
        and
        A.R. Bishop$^a$}

\begin{document}

\begin{abstract}
We argue that the superconducting state found in high-$T_c$ cuprates is
inhomogeneous with a corresponding inhomogeneous superfluid density. We
introduce two classes of microscopic models which capture the magnetic
and superconducting properties of these strongly correlated materials.
We start from a generalized $t$-$J$ model, in which appropriate
inhomogeneous terms mimic stripes. We find that inhomogeneous
interactions that break magnetic symmetries are essential to induce
substantial pair binding of holes in the thermodynamic limit. We argue
that this type of model reproduces the ARPES and neutron scattering
data seen experimentally. 
\vspace{1pc}
\end{abstract}

\maketitle

\section{Introduction and Scenario}

Few recent problems in science have generated so many controversial
discussions as the problem of high temperature superconductivity since
its experimental discovery in 1986. Two fundamental questions are: {\bf
Is the superconducting state found in high-$T_c$ cuprates homogeneous}?
{\bf Is the superfluid density in these materials homogeneous}? The
standard approach consists in accepting a homogeneous superconducting
state of various forms. We assert that this state is in fact
``inhomogeneous.'' In this work we summarize \cite{we} and expand on a
new scenario for striped superconductors where the interplay between
inhomogeneous superfluid density and phase  fluctuations determines the
critical temperature. 

In their undoped state, cuprates behave as antiferromagnetic (AF) Mott
insulators and it is precisely upon doping with holes that these
strongly correlated materials become superconductors. Recent
experiments seem to indicate that inhomogeneously textured
(intrinsically nanoscale) phases characterize the quantum state of high
temperature superconductors. This is, probably, not surprising in
retrospect since these are complex materials with competing time and
length scales arising from different interactions. A relevant and
non-trivial question is, however, whether  those textures are essential
to drive the phase coherent state, i.e., a Meissner phase. 

Neutron scattering experiments have proven to be a very useful tool in
investigating magnetic and superconducting properties of high-$T_c$
cuprate oxides. With improved sample quality and  resolution there is
reliable evidence for an incommensurate structure in the spin
susceptibility. On the other hand, recent angle-resolved photoemission
spectroscopy (ARPES) data suggest a one-dimensional (1D) like
electronic structure consistent with clustering of charge carriers into
1D channels \cite{expe}. Therefore, although the orientation, width,
length and dynamics of the channels remains to be elucidated, both the
above classes of experiments appear to confirm a new paradigm of spin and
charge ordering in high-$T_c$ superconductors: the ``stripe'' phase. 

Motivated by this new paradigm we introduced \cite{we} a class of
inhomogeneous microscopic models which capture the magnetic and
superconducting properties of these strongly correlated materials. The
origin(s) of the mesoscopic skeleton of stripe segments in the CuO$_2$
planes is presently unclear and several mechanisms could be
responsible, such as local spin-orbit coupling, Jahn-Teller
distortions, oxygen buckling at the stripe, and/or other magnetoelastic
effects. (Local  charge-lattice coupling may be an important source of
texture formation.) We showed, in particular, that appropriate
inhomogeneous interactions that break magnetic symmetries are
distinctive in inducing substantial pair binding of holes, as well as
explaining the magnetic neutron scattering properties. Moreover, based
upon the phenomenology of our microscopic model we developed a
mean-field (``Josephson spaghetti'') model which provides a scenario
for the macroscopic superconducting state. We also discussed the
connection of the resulting inhomogeneity-induced superconductivity to
recent experimental evidence for a linear relation between magnetic
incommensurability and the superconducting transition temperature, as a
function of doping. In a different work \cite{we1} we studied the
spectral properties of these inhomogeneous models and found, consistent
with experiments \cite{expe}, a flat band and the correct distribution
of quasiparticle weights.

In previous work \cite{we,we1} we have assumed static magnetic
inhomogeneities. Certainly, the stripe segments in real materials are
likely to have an intrinsic dynamics on a characteristic time scale
$\tau$. We assume that this time is large enough for attractive forces
to produce bound states of two holes. On the other hand, this stripe
dynamics will probably restore the $SU(2)$ spin rotation invariance on
time scales greater than $\tau$. $SU(2)$ symmetry does not have to be
broken statically. In the present manuscript we also discuss an
alternative approach where the magnetic inhomogeneities follow the
hole, i.e., the stripe phase is dynamically generated by the holes.
This model we call the {\it selfconsistent perturbing hole} (SPH) model
since the hole itself carries the perturbation. The main qualitative
difference between both classes of models is the absence of a broken
lattice translational symmetry state in the SPH case. Otherwise, the
basic phenomenology is qualitatively the same:  Holes pair in stripes
as a consequence of the existence of an AF background (avoiding a
possible global phase separation). The pairing mechanism is kinetic
exchange-interaction based and is provided by magnetic inhomogeneities
that {\it locally} break spin-rotational invariance. In the superfluid
phase, it is argued that a phase-locked state is generated as a
consequence of a coherent Josephson tunneling of the hole pairs between
and along stripes \cite{we}. 

\section{Microscopic Inhomogeneous Models}

In this Section we will present two classes of inhomogeneous models. 
The first model was already introduced in Ref. \cite{we}, where the
basic microscopic scenario starts from a homogeneous $t$-$J$
Hamiltonian as background 
\begin{equation} 
\label{H} 
H_{t\!-\!J} = -t  \sum_{\langle {\bf r,\bar{r}} \rangle, \sigma}
c^{\dagger}_{{\bf r} \sigma} c^{\;}_{{\bf \bar{r}} \sigma} + J \sum_{\langle
{\bf r,\bar{r}} \rangle} ({\bf S}_{\bf r} \cdot {\bf S}_{{\bf \bar{r}}} -
\frac{1}{4} \bar{n}_{\bf r} \bar{n}_{{\bf \bar{r}}} )  
\end{equation} 
but, to mimic the stripe segments, we add inhomogeneous magnetic
interactions. These inhomogeneous terms break translational invariance
and spin-rotational $SU(2)$ symmetry locally: 
\begin{equation} 
H_{\rm inh} = \sum_{\langle \alpha,\beta \rangle} \delta J_z \
S^z_\alpha S^z_\beta + \frac{\delta J_{\perp}}{2} \left( S^+_\alpha 
S^-_\beta + S^-_\alpha S^+_\beta \right)
\nonumber  
\label{H-inh} 
\end{equation} 
with $\delta J_{\perp} \neq \delta J_z$, representing the magnetic 
perturbation of a static local {\it Ising} anisotropy, locally lowering
spin symmetry (named a $t$-$JJ_z$ model). Only a few links (where the
stripes are located) have this lowered spin symmetry. This Ising
anisotropy is also sufficient to produce a spin gap. 

Pair binding of holes in this class of models is substantial \cite{we}.
This substantial binding energy is achieved as the energy for two
holes falls much faster with $t$ than twice the energy of one hole. 

It is interesting to make the following remarks: Suppose that a stripe
segment is represented by the 1D $t$-$J_z$ model
\begin{equation} 
\label{H1} 
H = -t  \sum_{\alpha, \sigma} (c^{\dagger}_{\alpha \sigma}
c^{\;}_{\alpha+1 \sigma} + {\rm H.c.}) + J_z \sum_{\alpha} S^z_{\alpha}
S^z_{\alpha+1} .
\end{equation} 
One can show \cite{cristian} that, within the ground state subspace
(for a given number of holes), this Hamiltonian maps into the
attractive spinless fermion model
\begin{equation} 
H = -t  \sum_{\alpha} (b^{\dagger}_{\alpha} b^{\;}_{\alpha+1} +{\rm
H.c.}) - \frac{J_z}{4} \sum_{\alpha} \tilde{n}_{\alpha} ,
\tilde{n}_{\alpha+1} 
\end{equation}
where $\tilde{n}_{\alpha}=b^{\dagger}_{\alpha} b^{\;}_{\alpha}$, and
which certainly has a superconducting phase (i.e., correlation exponent
$K_{\rho} > 1$) \cite{ours}. In this particular model, it turns out
that (e.g., at half-filling for $|J_z/8t| < 1$) the isolated stripe
segment belongs to the Luttinger liquid universality class. Notice,
however, that our stripes are embedded in an AF background. This
background provides a strong boundary condition that results in an
additional attractive potential for the holes in the stripe. As a
result, an enhanced superconducting region is expected \cite{cristian},
avoiding alternative charge density wave phases or phase segregation. 

%
%

Knowing that perturbating the system by breaking magnetic symmetries is
an efficient mechanism to achieve substantial binding of carriers, some
natural questions arise: What would happen if the hole itself carries
this perturbation? Would this process be sufficient to generate a
stripe phase? Would the binding energy be still appreciable in the
thermodynamic limit? 

\begin{figure}[tbh]
\epsfverbosetrue
\epsfxsize=5.5cm
\vspace*{-0.5cm}
\centerline{ \rotate[r] {\epsfbox{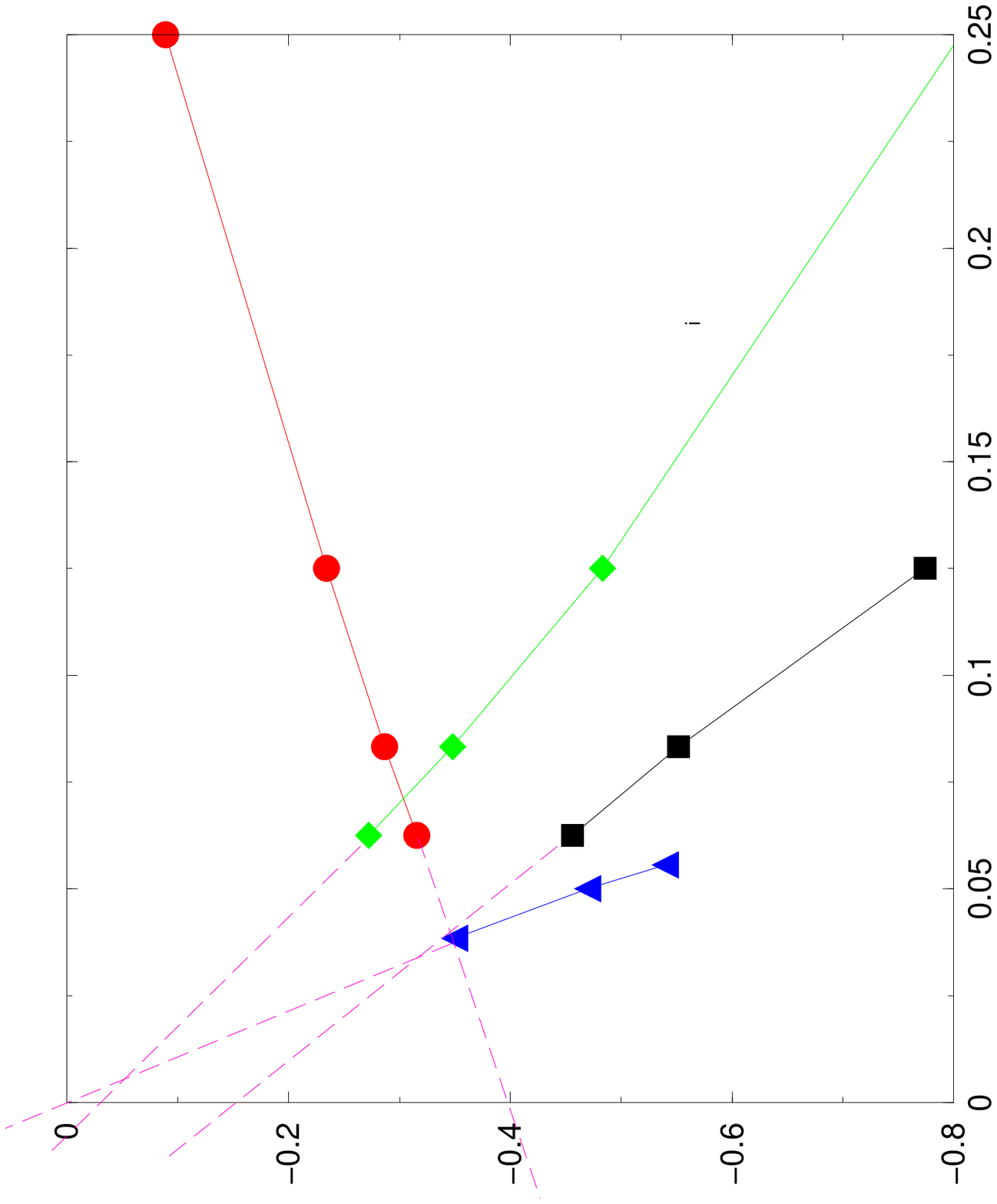}}
\put(-55.0,105){{\normalsize{ $t$-$JJ_z$ }}}
\put(-55.0,40){{\normalsize{ $t$-$J$ 1D }}}
\put(-173,50){{\normalsize{ $t$-$J$ 2D }}}
\put(-155,20){{\normalsize{ $t$-$J$ + $H_{\rm sph}$ }}}
\put(-40,-10){{\normalsize{ $1/N$ }}}
\put(-205,100){\rotate[l]{\normalsize{ ${E_b}$ }}}
}
\caption{Binding energy $E_b$ as a function of $1/N$ for different
models.  The lines are guides to the eye. Both homogeneous $t$-$J$
models in 1D (diamonds) and 2D (triangles) ($t=J=1$) extrapolate to
zero or  negligible binding energy. $t$-$JJ_z$  extrapolates to a huge
binding energy while $t$-$J$ + $H_{\rm sph}$ (in both $\delta
J_{\perp}=-0.9$) extrapolates to moderate binding energy.}
\label{scaling}
\end{figure}

Our SPH model corresponds to the Hamiltonian $H=H_{t\!-\!J}+H_{\rm
sph}$, where

\begin{equation}
H_{\rm sph}=\sum_{{\bf r,d}} \frac{\delta J_{\perp}}{2} (1 -
\bar{n}_{\bf r}) \left( S^+_{{\bf r}}  S^-_{{\bf r}+{\bf d}} +
S^-_{\bf r} S^+_{{\bf r}+{\bf d}} \right) .
\label{Hph}
\end{equation}
In this Hamiltonian $1 - \bar{n}_{\bf r} = n_{\bf r}$ is the ocupation
number of holes at site ${\bf r}$ and $S^+_{{\bf r}}$, $S^-_{{\bf r}}$
are the usual spin operators \cite{we}. The presence of a hole at site
${\bf r}$ perturbs the magnetic links in the directions defined by
${\bf d}$ by lowering the spin symmetry and making them more
Ising-like.

This model is perhaps more natural on physical grounds than the
$t$-$JJ_z$ one. Magnetoelastic effects caused by the presence of the
hole, or buckling of the oxygens close to the carrier may easily
produce an Ising-like anisotropy. Upon doping with holes it is not
obvious what the extra hole does to the environment. Since these are
strongly correlated materials, the extra holes could have a stronger
influence on the system than just those effects produced by simple
hopping dynamics.

The models considered above are different in some respects. While in
the model of Eqs. \ref{H},\ref{H-inh} translational symmetry has been
explicitly broken (i.e., adding  $H_{\rm inh}$) the model defined by
$H_{\rm sph}$ is translationally symmetric, and the only symmetry that
has been explicitly broken is the spin $SU(2)$ symmetry around each
hole. This fact has some direct experimental consequences. As already
discussed in Ref. \cite{we}, the $t$-$JJ_z$ model has an inhomogeneous
hole density,  as they prefer to occupy sites where the magnetic links
have been weakened (the stripes). This is not the case for the SPH
model since, as the translational symmetry has not been broken, the
holes will be found with equal probability on every lattice site. This
fundamental difference could be resolved by a Scanning Tunneling
Microscope experiment.
 
In this work we will not address the issue of stripe formation in the
SPH model, but rather concentrate on the hole pairing in those
textures. Clearly there will be a competition between kinetic and
magnetic energies. While the first will try to delocalize the pair, the
former will contribute to the pairing. In Fig. \ref{scaling} we show
the binding energy (defined as $E_b=E_2+E_0-2 E_1$, where $E_i$ is the
ground state energy in the subspace with $i$ holes) as a function of
$1/N$ (where $N$ is the size of the system). $t$-$J$ models in 1D
(diamonds) and 2D (triangles) have zero or negligible binding in the
extrapolated thermodynamic limit. The model labeled as $t$-$JJ_z$
corresponding to the Hamiltonian of Eqs. \ref{H} ,\ref{H-inh} in 1D has
substantial binding in the thermodynamic limit. 
\begin{figure}[htb]
\epsfverbosetrue
\epsfxsize=10cm
\vspace*{-0.01cm}
\centerline{ \rotate[r] {\epsfbox{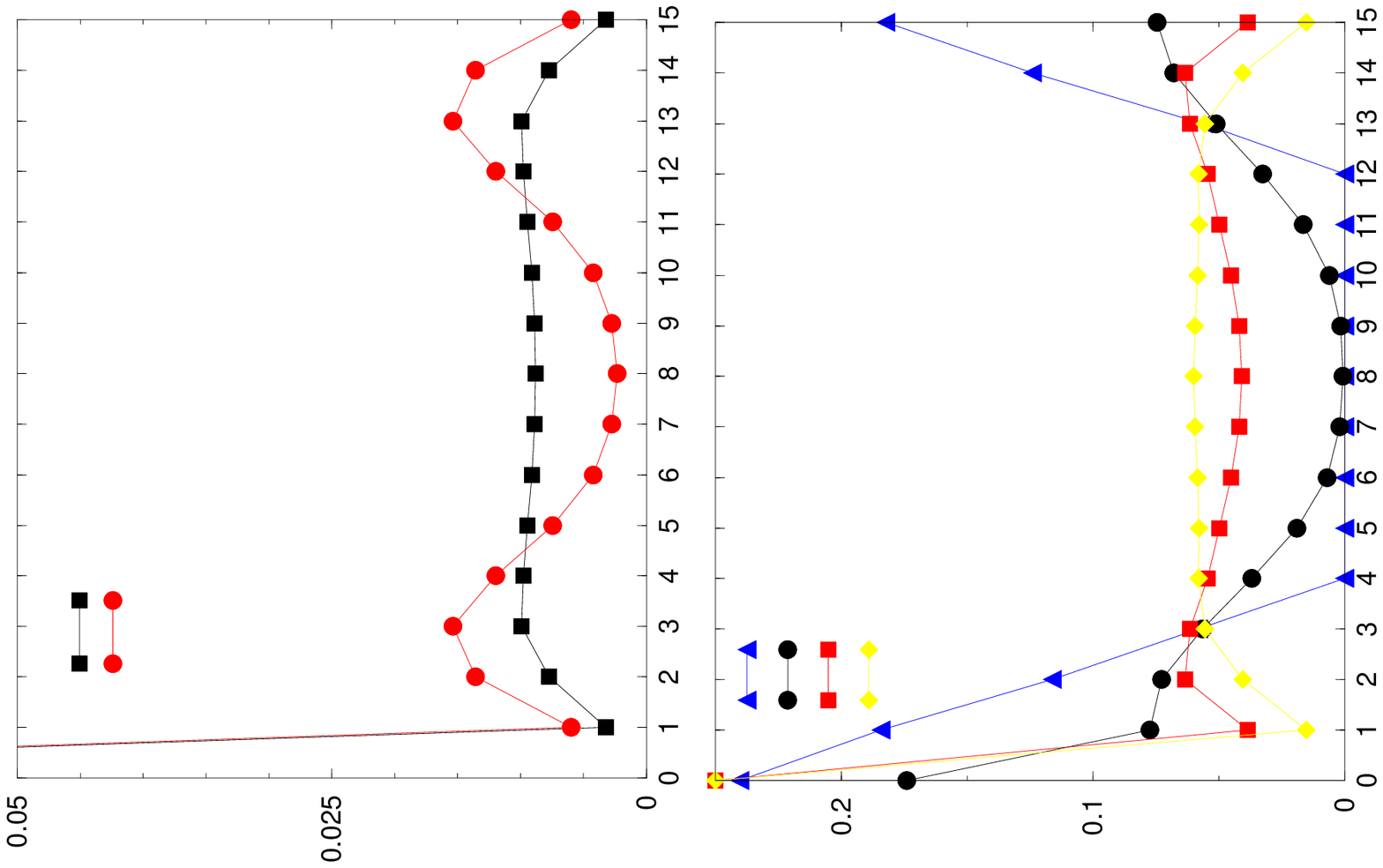}}
\put(-120,262){{\scriptsize{{\bf d}=2 }}}
\put(-120,255){{\scriptsize{{\bf d}=3}}}
\put(-129,126){{\scriptsize{ $t=0.05$ }}}
\put(-129,118){{\scriptsize{ $t=0.25$ }}}
\put(-129,110){{\scriptsize{ $t=0.5$ }}}
\put(-129,102){{\scriptsize{ $t=1.5$ }}}
\put(-40,-12){{\normalsize{ site i}}}
\put(-190,70){\rotate[l]{\normalsize{ ${\bf \langle n_0 \cdot n_i \rangle}$ }}}
\put(-190,200){\rotate[l]{\normalsize{ ${\bf \langle n_0 \cdot n_i \rangle}$ }}}
}
\caption{The correlation function $\langle n_0 \cdot n_i
\rangle$, which reflects the shape of the pair, for different cases.
Upper panel shows the cases where the perturbation carried by the hole
extends to $d=2$ or 3. Lower panel shows the same correlation
function for different values of the kinetic energy $t$. The pair
evolves from a tightly bound state to an extended one as the result of a
competition between the kinetic energy and the magnetism.}
\label{noni}
\end{figure}
That model corresponds to  placing a perturbation like Eq. \ref{H-inh}
every 4 sites. The model labeled as $t$-$J$ + $H_{\rm sph}$ is the one
we are mostly interested in this work. It has appreciable binding of
holes in the thermodynamic limit, although not as strong  as in the
$t$-$JJ_z$ model. In the calculations we have taken $|{\bf d}|=2$
(i.e., the hole perturbs up to second neighbors). For $|{\bf d}|=1$ the
model reduces trivially to the $t$-$J$ model, as the links close to the
holes are magnetically inactive. Therefore $|{\bf d}|=d$ must be
greater or equal to 2 to show any new behavior. In order to get
information about the bound state we have calculated the correlation
function $\langle n_0 \cdot n_i \rangle$, which gives the probability
of finding a hole at site $i$ if there is one at site $0$. We remark
again that the density of holes is constant and equal to $\langle n_i
\rangle=N_h/N$ (with $N_h$ the number of holes and $N$ the size of the
system) for every site $i$ in the case of $t$-$J$ + $H_{\rm sph}$. In
Fig. \ref{noni}, upper panel we show this correlation function for the
cases $d=2$ and $d=3$ when there are two holes in a 16 sites chain. As
expected, the pair is more strongly bound for $d=3$. In a real material
there must be a decreasing function of the perturbation carried by the
hole with distance. In our case that perturbation is constant. In
Fig. \ref{noni} lower panel, we show the same correlation function for
$d=2$ and different $t$ values. The pair evolves from being tightly
bound to delocalization as the kinetic energy is increased. It is worth
noting that even when the holes are not very close the binding energy
scales to a finite value.

We are grateful to C.D. Batista for useful discussions. This work has
been supported by the US DOE.

\end{document}